# A Quantum Mechanical Approach for the Simulation of Si/SiO$_2$ Interface Roughness Scattering in Silicon Nanowire Transistors

JING WANG, ERIC POLIZZI*, AVIK GHOSH, SUPRIYO DATTA, AND MARK LUNDSTROM

*School of Electrical and Computer Engineering, *Department of Computer Sciences, Purdue University, West Lafayette, Indiana 47907, U. S. A.*

*jingw@purdue.edu*

**Abstract.** In this work, we present a quantum mechanical approach for the simulation of Si/SiO$_2$ interface roughness scattering in silicon nanowire transistors (SNWTs). The simulation domain is discretized with a three-dimensional (3D) finite element mesh, and the microscopic structure of the Si/SiO$_2$ interface roughness is directly implemented. The 3D Schrödinger equation with open boundary conditions is solved by the non-equilibrium Green's function method together with the coupled mode space approach. The 3D electrostatics in the device is rigorously treated by solving a 3D Poisson equation with the finite element method. Although we mainly focus on computational techniques in this paper, the physics of SRS in SNWTs and its impact on the device characteristics are also briefly discussed.

## 1. Introduction

With the rapid progress in nanofabrication technology, semiconducting nanowires have been extensively studied for the potential applications on nanoelectronics. The Silicon Nanowire Transistor (SNWT), in particular, has attracted broad attention as a promising device structure for future integrated circuits [1, 2]. As a result, the exploration of carrier transport and the modeling of various scattering mechanisms in Si nanowires have become increasingly important.

It is well known that transport in ultra-thin body MOSFETs is greatly degraded by the Si/SiO$_2$ interface roughness scattering (so called "surface roughness scattering" (SRS) [3, 4]). In an SNWT, the channel is surrounded by the Si/SiO$_2$ interfaces and one may expect strong SRS to be observed in SNWTs. However, recent experiments show an unexpectedly high mobility in Si nanowires [1]. To fully understand carrier transport in Si nanowires, a careful theoretical study, particularly a detailed simulation of SRS in SNWTs, is essentially important.

The effects of Si/SiO$_2$ interface roughness on carrier transport are as follows: 1) due to the dielectric constant difference between Si and SiO$_2$, the roughness introduces electrostatic potential variations inside the Si body, which behave as a scattering potential for carriers, 2) due to the Si/SiO$_2$ conduction band-edge discontinuity, the roughness causes a fluctuating electron subband energy and wavefunction shape, which lowers the transmission from the source to the drain (so called "wavefunction deformation scattering" [5, 6]). When the device size is relatively large and quantum confinement is weak, the first effect dominates, and SRS is well described by the first order perturbation theory and semiclassical models such as the Monte-Carlo approach [3, 4]. In SNWTs with very small diameters (e.g., <5nm), both effects become substantially important and a direct treatment of SRS in a quantum mechanical framework is necessary.

In this work, we present a quantum mechanical approach for the simulation of SRS in SNWTs. A three-dimensional (3D) finite element [7] mesh is adopted to discretize the simulation domain, and the microscopic structure of the Si/SiO$_2$ interface roughness is directly implemented. The 3D Schrödinger equation with open boundary conditions is solved by the non-equilibrium Green's function (NEGF) method [8] together with the coupled mode space approach [7]. By using the finite element method, the 3D Poisson equation is self-consistently solved with the transport equation. In the following sections, we will discuss the details of the implementation of Si/SiO$_2$ interface roughness (section 2) and the coupled mode space approach used to obtain the device Hamiltonian for the NEGF calculation (section 3). Although we mainly focus on computational techniques in this paper, the physics of SRS in SNWTs and its impact on the device characteristics will also be briefly discussed (section 4).

## 2. Device Structure and Implementation of Si/SiO$_2$ Interface Roughness

Figure 1 (a) shows the simulated device structure in this work. It is a gate-all-around SNWT with a rectangular cross-section and a transport direction along the [100] axis. The source/drain (S/D) doping concentration is $2\times10^{20} cm^{-3}$ and the channel is undoped. There is no S/D overlap with channel and the gate length is equal to the channel length ($L=10$nm). For the nominal device (with smooth Si/SiO$_2$ interfaces), the Si body thickness and width are both 3nm and the oxide thickness is 1nm.

As mentioned earlier, the simulation domain is discretized by a 3D finite element mesh and each element is a triangular prism with a *2Å* height and edge length, comparable to the size of roughness at the (100) Si/SiO$_2$ interfaces [9]. To implement the Si/SiO$_2$ interface roughness, a 2D random distribution is generated across the *whole* Si/SiO$_2$ interface (unfolding the four interfaces into a rectangle) according to an exponential auto-covariance function [9],

$$C(n) = \Delta_m^2 e^{-\sqrt{2}n\Delta x/L_m} \quad (1)$$

where $\Delta x=1.92Å$ is the sampling interval, $L_m=7.1Å$ is the correlation length, and $\Delta_m=1.4Å$ is the rms fluctuation of the roughness. Based on this random distribution, the (material) types of the elements (prisms) at the Si/SiO$_2$ interfaces may be changed from Si to SiO$_2$ or reversely, to mimic the roughened interfaces, as shown in Fig. 1(b).

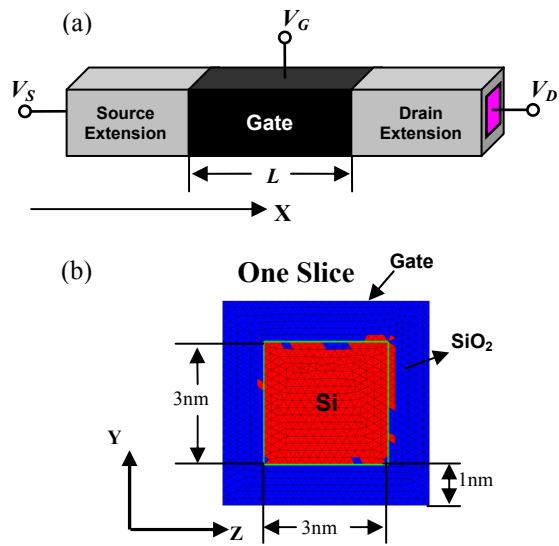

*Figure 1*. (a) The schematic diagram of the simulated gate-all-round SNWT with a rectangular cross-section ($L=10$nm). The X, Y, and Z coordinates are in the [100], [010] and [001] crystal orientations, respectively. $V_S$, $V_D$, $V_G$ are the applied voltage biases on the source, drain and gate. (b) A slice of the SNWT with a rough Si/SiO$_2$ interface (please note that the pattern of the roughness varies from slice to slice).

## 3. Coupled Mode Space Approach and NEGF

After the Si/SiO$_2$ roughness is implemented, the roughened SNWT is simulated by the NEGF approach. As the first step of the NEGF calculation, a proper representation of the 3D device Hamiltonian, $H_{3D}$, needs to be obtained. A 3D real space discretization of $H_{3D}$ would result in a ~130,000×130,000 matrix for the simulated structure in this work, which would cause a huge computational burden. In our work, instead of using the 3D real space representation, we adopt a coupled mode space (CMS) approach [7], which provides accurate results as compared with the 3D real space approach while significantly reducing the computational complexity. Here we briefly discuss the basics of the CMS approach, and more details can be found in [7].

The 3D stationery Schrödinger equation is written as

$$H_{3D}\Psi(x,y,z) = E\Psi(x,y,z). \quad (2)$$

We adopt the effective-mass approximation and the device Hamiltonian is expressed as

$$H_{3D} = -\frac{\hbar^2}{2m_x^*(y,z)}\frac{\partial^2}{\partial x^2} - \frac{\hbar^2}{2}\frac{\partial}{\partial y}\left(\frac{1}{m_y^*(y,z)}\frac{\partial}{\partial y}\right),$$
$$-\frac{\hbar^2}{2}\frac{\partial}{\partial z}\left(\frac{1}{m_z^*(y,z)}\frac{\partial}{\partial z}\right) + U(x,y,z) \quad (3)$$

where $m_x^*$, $m_y^*$ and $m_z^*$ are the electron effective masses in the $x$, $y$, and $z$ directions, respectively, and $U(x,y,z)$ is the electron conduction band-edge profile in the active device. It should be noted that the penetration of electron wavefunction into the oxide region is considered in our simulation, which is required for the effective-mass approximation to be valid for Si nanowire simulations [10].

Expanding the 3D wavefunction, $\Psi(x,y,z)$, in the 2D eigenfunction space, we obtain

$$\Psi(x,y,z) = \sum_n \varphi^n(x) \cdot \xi^n(y,z;x), \quad (4)$$

where $\xi^n(y,z;x)$ is the $n^{th}$ eigenfunction of the following 2D Schrödinger equation in the cross-sectional plane,

$$\left[-\frac{\hbar^2}{2}\frac{\partial}{\partial y}\left(\frac{1}{m_y^*(y,z)}\frac{\partial}{\partial y}\right) - \frac{\hbar^2}{2}\frac{\partial}{\partial z}\left(\frac{1}{m_z^*(y,z)}\frac{\partial}{\partial z}\right)\right.$$
$$\left. + U(x_0,y,z)\right] \cdot \xi^n(y,z;x_0) = E_{sub}^n(x_0) \cdot \xi^n(y,z;x_0) \quad (5)$$

where $E_{sub}^n(y,z;x_0)$ is the $n^{th}$ subband level at $x=x_0$. The eigenfunctions, $\xi^n(y,z;x)$ ($n=1, 2,...$) satisfy the orthogonal condition for any $x$,

$$\iint_{y,z} \xi^m(y,z;x)\xi^n(y,z;x)dydz = \delta_{m,n}, \quad (6)$$

where $\delta_{m,n}$ is the Kronecker delta function.

Inserting (4) into (3), multiplying both sides by $\xi^m(y,z;x)$ and then doing an integral over the $y$-$z$ plane, a coupled 1D Schrödinger equation is obtained,

$$-\frac{\hbar^2}{2}\left(\sum_n a_{mn}(x)\right)\frac{\partial^2}{\partial x^2}\varphi^m(x) - \frac{\hbar^2}{2}\sum_n c_{mn}(x)\varphi^n(x)$$
$$-\hbar^2\sum_n b_{mn}(x)\frac{\partial}{\partial x}\varphi^n(x) + E_{sub}^m(x)\varphi^m(x) = E\varphi^m(x) \quad (7)$$

where

$$a_{mn}(x) = \iint_{y,z} \frac{1}{m_x^*(y,z)}\xi^m(y,z;x)\xi^n(y,z;x)dydz, \quad (8)$$

$$b_{mn}(x) = \iint_{y,z} \frac{1}{m_x^*(y,z)}\xi^m(y,z;x)\frac{\partial}{\partial x}\xi^n(y,z;x)dydz, \quad (9)$$

and

$$c_{mn}(x) = \iint_{y,z} \frac{1}{m_x^*(y,z)}\xi^m(y,z;x)\frac{\partial^2}{\partial x^2}\xi^n(y,z;x)dydz. \quad (10)$$

For SNWTs, the electron wavefunction is well confined in the Si region, therefore, we can neglect $a_{mn}$ if $m \neq n$ and simplify (7) as

$$-\frac{\hbar^2}{2}a_{mm}(x)\frac{\partial^2}{\partial x^2}\varphi^m(x) - \frac{\hbar^2}{2}\sum_n c_{mn}(x)\varphi^n(x)$$
$$-\hbar^2\sum_n b_{mn}(x)\frac{\partial}{\partial x}\varphi^n(x) + E_{sub}^m(x)\varphi^m(x) = E\varphi^m(x) \quad (11)$$

From the derivation above, it is clear that the CMS approach is mathematically equivalent to the 3D real space approach if all the confined modes are included (i.e., $m, n=1, 2, …, N_{YZ}$, where $N_{YZ}$ is the number of the nodes in the $y$-$z$ plane). In SNWTs with small diameters, however, only a few lowest subbands are occupied (i.e., $m, n=1, 2, …, M, M \ll N_{YZ}$) due to strong quantum confinement. For this reason, we can start with $M=1$ in our calculation, and then increase $M$ one by one until all the device characteristics (e.g., electron density and current) saturate. For the simulated structure in this work (see Fig. 1), it is found that $M=2$ is sufficient for each of the six valleys.

Based on (11), the device Hamiltonian is obtained as,

$$H\begin{bmatrix}\varphi^1\\\varphi^2\\...\\...\\\varphi^M\end{bmatrix} = \begin{bmatrix}h_{11} & h_{12} & h_{13} & ... & h_{1M}\\h_{21} & h_{22} & h_{23} & ... & h_{2M}\\... & ... & ... & ... & ...\\... & ... & ... & ... & ...\\h_{M1} & h_{M2} & h_{M3} & ... & h_{MM}\end{bmatrix}\begin{bmatrix}\varphi^1\\\varphi^2\\...\\...\\\varphi^M\end{bmatrix} = E\begin{bmatrix}\varphi^1\\\varphi^2\\...\\...\\\varphi^M\end{bmatrix}, \quad (12)$$

where

$$h_{mn} = \delta_{m,n}\left[-\frac{\hbar^2}{2}a_{mm}(x)\frac{\partial^2}{\partial x^2} + E_{sub}^m(x)\right] - \frac{\hbar^2}{2}c_{mn}(x) - \hbar^2 b_{mn}(x)\frac{\partial}{\partial x},$$
$$(m,n=1,2,...,M). \quad (13)$$

As discussed in [7], $b_{mn}$ and $c_{mn}$ represent the coupling between different modes. For SNWTs with *smooth* interfaces, $b_{mn}$ and $c_{mn}$ become negligible and $h_{mn}=0$ when $m \neq n$, so the uncoupled mode space approach can be adopted [7]. When the Si/SiO$_2$ interface roughness is present, however, the coupling terms, $b_{mn}$ and $c_{mn}$, are not negligible anymore. So the simulation of roughened SNWTs is a good example for the application of the CMS approach to quantum transport.

After the device Hamiltonian is obtained, the electron density and current are calculated by the NEGF approach. To be concise, we refer the readers to [7, 8] for more details.

## 4. Results and Conclusions

In our simulation, to emphasize the role of SRS on electron transport, we do not include any other scattering mechanisms, so electron transport is coherent inside the device. The simulation results for the roughened device are then compared with those for a structure with the same geometrical parameters but smooth Si/SiO$_2$ interfaces, which is treated as the *ballistic* device.

Figure 2 illustrates the effects of SRS on both (a) the internal parameters (e.g., subbands, transmission) and (b) current-voltage characteristics of the simulated SNWT. It is found that 1) SRS significantly deforms the electron subbands in the SNWT, 2) SRS blocks low-energy injections from the source into the channel, so it reduces the density-of-states (DOS) at low energy and consequently raises the threshold voltage of the device, 3) due to the DOS degradation caused by SRS, the roughened SNWT displays lower subband levels at the top of the barrier under ON-state conditions, which somehow compensates the reduction of transmission due to SRS at the ON-state, and 4) the transmission reduction caused by SRS becomes quite significant when more than one subbands become conductive (this is consistent with the observation that SRS is very important in planar MOSFETs, where a large number of transverse modes are conductive). In conclusion, this work provides an opportunity to understand the physics of SRS in SNWTs, which can be substantially different from that in planar MOSFETs.

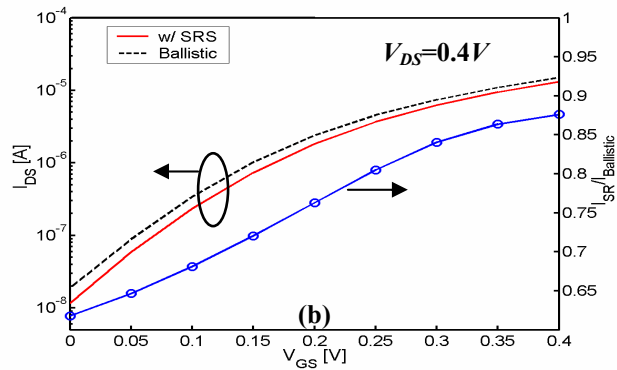

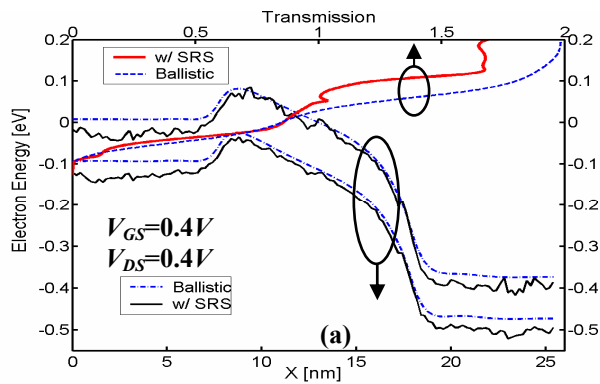

*Figure 2.* (a) The electron subband and transmission for the simulated SNWT with and without (ballistic) SRS. (b) The *I-V* characteristics (left) of the simulated SNWT and the ratio (right) of the current for the roughened SNWT, $I_{SR}$, to the ballistic current, $I_{Ballistic}$.

## Acknowledgements

This work is supported by Semiconductor Research Corporation (SRC), the MARCO focus center on Materials, Structures and Devices (MSD) and NSF Network on Computational Nanotechnology (NCN).